\documentclass[prb,twocolumn,showpacs,amsmath,amssymb,superscriptaddress]{revtex4}
\usepackage[dvips]{graphicx}
\usepackage{graphics}
\usepackage{dcolumn}
\usepackage{bm}

\begin{document}
\title{Low-temperature magnetization of  (Ga,Mn)As
semiconductors}
\author{T.~Jungwirth}
\affiliation{Institute of Physics  ASCR, Cukrovarnick\'a 10, 162 53 Praha 6, Czech Republic }
\affiliation{School of Physics and Astronomy, University of Nottingham, Nottingham NG7 2RD, UK}
\author{J.~Ma\v{s}ek}
\affiliation{Institute of Physics  ASCR, Na Slovance 2, 182 21 Praha 8, Czech Republic }
\author{K.Y.~Wang}
\affiliation{School of Physics and Astronomy, University of Nottingham, Nottingham NG7 2RD, UK}
\author{K.W.~Edmonds}
\affiliation{School of Physics and Astronomy, University of Nottingham, Nottingham NG7 2RD, UK}
\author{M.~Sawicki}
\affiliation{Institute of Physics, Polish Academy of Sciences, 02668 Warszawa, Poland}
\author{M.~Polini}
\affiliation{NEST-INFM and Scuola Normale Superiore, I-56126 Pisa, Italy}
\author{Jairo~Sinova}
\affiliation{Department of Physics, Texas A\&M University, College Station,Texas 77843-4242, USA}
\author{A.H.~MacDonald}
\affiliation{Department of Physics, University of Texas at Austin, Austin TX 78712-1081}
\author{R.P.~Campion}
\author{L.X.~Zhao}
\author{N.R.S.~Farley}
\affiliation{School of Physics and Astronomy, University of Nottingham, Nottingham NG7 2RD, United Kingdom}
\affiliation{CCLRC Daresbury Laboratory, Warrington WA4 4AD, United Kingdom}
\author{T.K.~Johal}
\author{G.~van der Laan}
\affiliation{CCLRC Daresbury Laboratory, Warrington WA4 4AD, United Kingdom}
\author{C.T.~Foxon}
\author{B.L.~Gallagher}
\affiliation{School of Physics and Astronomy, University of Nottingham, Nottingham NG7 2RD, UK}
\begin{abstract}
We report on a comprehensive study of the ferromagnetic
moment per Mn atom in (Ga,Mn)As ferromagnetic semiconductors. Theoretical discussion
is based on microscopic calculations and on an effective model of Mn local moments antiferromagnetically coupled to valence band hole spins.
The validity of the effective model over the range of doping studied is assessed by comparing with microscopic 
tight-binding/coherent-potential approximation calculations. 
Using the virtual crystal ${\protect\rm k}\cdot {\protect\rm p}$ model for 
hole states, we evaluate the zero-temperature mean-field contributions to the magnetization
from the hole kinetic and exchange energies, and magnetization suppression due to 
quantum fluctuations of Mn moment orientations around
their mean-field ground state values. Experimental low-temperature ferromagnetic moments per
Mn are obtained by superconducting quantum interference device
and x-ray magnetic circular dichroism measurements in a series of  (Ga,Mn)As semiconductors
with nominal Mn doping ranging from $\sim$2\% to 8\%.
Hall measurements in as-grown and annealed samples are used to estimate the number
of uncompensated substitutional Mn moments. Based on our comparison between experiment and theory 
we conclude that all these Mn moments in high quality (Ga,Mn)As materials have nearly parallel ground state 
alignment.  
\end{abstract}

\maketitle

\section{Introduction}
Early experimental studies of (Ga,Mn)As ferromagnetic semiconductors, reporting large apparent 
magnetization deficits,\cite{Ohno:2001_a,Potashnik:2002_a}
motivated a theoretical search for possible intrinsic 
origins of frustrating magnetic interactions in this material. Using a wide spectrum
of computational techniques, ranging from {\em ab initio} LDA 
methods\cite{Mahadevan:2004_b,Kudrnovsky:2004_a} and microscopic 
tight-binding approximations\cite{Timm:2004_b} to semiphenomenological, 
${\protect\rm k}\cdot {\protect\rm p}$
kinetic-exchange 
models,\cite{Schliemann:2002_a,Schliemann:2003_a,Brey:2003_a,Zarand:2002_a,Fiete:2004_a}
the theoretical studies have identified several mechanisms that can 
lead to non-collinear ground states.  The observation that long wavelength 
spin-waves with negative energies frequently occur within
a spherically symmetric kinetic-exchange model illustrates\cite{Schliemann:2002_a} 
that randomness in the distribution
of Mn moments  can result in an instability of the collinear ferromagnetic state.
Frustration can be further enhanced when positional disorder is combined with anisotropies in 
Mn-Mn interactions. The {\em pd} character of electronic states forming the magnetic moment leads
to magnetic interaction anisotropies with respect to the crystallographic orientation of the 
vector connecting two Mn moments.\cite{Mahadevan:2004_b,Kudrnovsky:2004_a,Timm:2004_b,Brey:2003_a} 
When spin-orbit coupling is taken into 
account,\cite{Zarand:2002_a,Fiete:2004_a,Schliemann:2003_a,Timm:2004_b} magnetic interactions
also become anisotropic with respect to the relative orientation of the Mn-Mn connecting vector and the magnetic moment. 

Some degree of non-collinearity is inevitable as a combined consequence of positional disorder and 
spin-orbit coupling.  Nevertheless 
a large suppression of the ferromagnetic moment is not expected theoretically\cite{Timm:2004_b}
in metallic (Ga,Mn)As samples with Mn concentrations
above 1\%.  The minor role of non-collinearity is due 
largely to the long-range character of magnetic
interactions, which tends to average out the frustrating 
effect of anisotropic coupling between randomly distributed Mn impurities. 
In this paper we present detailed calculations of zero temperature magnetization in
(Ga,Mn)As ferromagnets and compare the results with superconducting quantum interference device
(SQUID) and x-ray magnetic circular dichroism (XMCD) measurements in a series
of samples with nominal Mn doping ranging from $\sim$2\% to 8\%.  Our calculations neglect 
effects that would lead to non-collinearity, appealing to expectations that these effects
are small. Our assumption is consistent with
experimental observations of larger ferromagnetic moments in recently synthesized high-quality
samples.\cite{Edmonds:2005_b,Jungwirth:2005_b} 
The substantial magnetization suppression seen in many early (Ga,Mn)As samples 
is attributed here primarily to the role played in those samples by 
interstitial Mn atoms.  
The consistency of the theoretical
and experimental data that we are able to achieve, allows us 
to  rule out any marked magnetic frustrations in the ground
state of high-quality (Ga,Mn)As ferromagnets, and helps to clarify the 
character of magnetic interactions in this material. 

 Two distinct theoretical approaches are used in the paper to discuss magnetization in (Ga,Mn)As semiconductors.
 In the more microscopic approach we account explicitly for the five $d$-orbital electrons on a substitutional
 Mn$_{\rm Ga}$ impurity, and for the strong on-site Coulomb correlations
 that suppress spin and charge fluctuations of the $L=0$, $S=5/2$ state of the atomic Mn $d$-shell. 
 Magnetism in  the mixed crystal
 arises in this picture from electron hopping between the Mn $d$-states and $p$-orbitals 
 concentrated on the As sublattice that form
 the top of the host semiconductor valence band. For
 weak $p-d$ hybridization, a second approach is possible.  The Schrieffer-Wolff transformation\cite{Schrieffer:1996_a} allows  us to
 map the microscopic Hamiltonian onto an effective Hamiltonian for local $S=5/2$ moments and valence 
 band states whose coupling is described by the kinetic-exchange term, 
 $J_{pd}\hat{\bf S}({\bf R})\cdot\hat{\bf s}({\bf r})\delta({\bf R}-{\bf r})$,
 where $\hat{\bf S}$ and $\hat{\bf s}$ are the local moment and valence band state spin operators, respectively.
 This approach will fail if the $p-d$ hybridization is too strong, but appears to be reliable for (Ga,Mn)As. 

The paper is organized as follows: In Section~\ref{d5+hole} we identify the key physical considerations
related to ground-state magnetization of (Ga,Mn)As ferromagnets by
focusing first on a single Mn($d^5$+hole) complex. We recall the connection between $p-d$ hybridization
and the antiferromagnetic kinetic-exchange coupling, digress on the sign of the hole contribution to
total moment per Mn, and discuss the expected mean-field contribution
to magnetization per Mn from the Mn local moment and from the antiferromagnetically
coupled hole. We also explain that quantum fluctuations around the mean-field
ground state are generically present because of antiferromagnetic character of the 
$p-d$ kinetic exchange interaction. 

Magnetization calculations for the many-Mn-impurity system are discussed
in Section~\ref{manyMn}.  The relevant considerations here parallel those that 
apply for isolated Mn($d^5$+hole) complexes, but differ in detail because of 
interactions between moments.  Zero temperature ferromagnetic moments per Mn are first studied
within the tight-binding/coherent-potential approximation (TBA/CPA) model.
The results of these microscopic calculations indicate that (Ga,Mn)As is in a 
weak $p-d$ hybridization regime
over the whole range of Mn concentrations that we study. 
These calculations help establish theoretically the validity 
of the effective kinetic-exchange model. 
The virtual crystal approximation and 
the ${\protect\rm k}\cdot {\protect\rm p}$ effective Hamiltonian are then used to evaluate  
contributions to the mean-field magnetization from hole
kinetic and exchange energies
and to confirm the expected weak role  
quantum fluctuations around the mean-field many-body ground state. 

Experimental SQUID and XMCD
data are presented in Section~\ref{experiment}. Partial concentrations of substitutional and
interstitial Mn impurities and the corresponding number of uncompensated
local moments are derived from the nominal Mn doping and from Hall measurements of the hole density in
as-grown and annealed samples.\cite{Jungwirth:2005_b} 
The collinearity of the ferromagnetic ground state in the (Ga,Mn)As materials
we study is tested by comparing experimental data with theoretical calculations. 
Section~\ref{conclusions} briefly summarizes main conclusions of the paper.  

\section{Magnetization of an isolated ${\rm\bf Mn}({\protect\bf d^5}$+hole) complex}
\label{d5+hole}
\subsection{$d^{5}$+hole picture}
\begin{figure}[h]
\includegraphics[width=2.0in,angle=0]{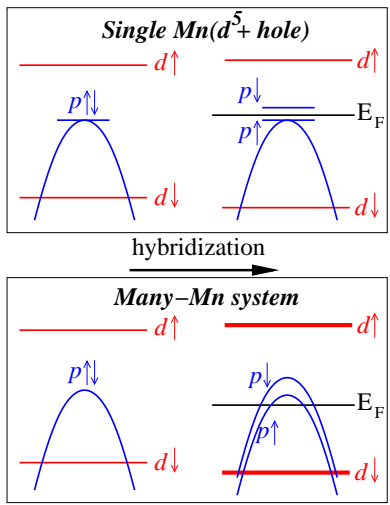}
\caption{Electron-picture cartoon: splitting of the isolated Mn acceptor level (top panel)
and of the top of the valence-band in the many-Mn system (bottom panel) due to p-d 
hybridization.} 
\label{hybrid}
\end{figure}

Most of the spectral weight near the top of the GaAs valence band originates from As 4$p$
levels.
Magnetic coupling between these states and the strongly localized 3$d$ electrons on
the substitutional Mn$_{\rm Ga}$ atom, which is at the origin
of ferromagnetism in (Ga,Mn)As materials, 
is dominated by the $p-d$ hybridization contribution.\cite{Bhattacharjee:1983_a} 
The majority spin $d$-shell, with all five electron spins aligned forming
a large spin $S=5/2$, has its spectral weight centered $\sim$ 3.5~eV below the
top of the valence band. The empty, minority-spin  $d$-level overlaps
with the conduction band.  Hybridization between $p$ and $d$ orbitals 
therefore, increases the energy of the majority-spin $p$ orbitals (level repulsion) and 
favors the occupation of minority-spin $p$ orbitals.  This is the basic origin of 
the antiferromagnetic interaction represented by $J_{pd}$, as explained further 
in Fig.~\ref{hybrid}.   

Apart from providing a magnetic moment, the substitutional Mn$_{\rm Ga}$ impurity
acts as a moderately shallow single-acceptor in GaAs.\cite{Linnarsson:1997_a}
The cartoon in top panel of Fig.~\ref{hybrid} shows the splitting of the impurity level
due to $p-d$ hybridization and the lower panel of Fig.~\ref{hybrid} illustrates
the many-Mn system in which the impurity level is broadened, eventually 
merging with the host valence band at higher Mn concentrations.  
To avoid confusion that may result from using the hole picture to describe magnetization
of carriers in p-type (Ga,Mn)As materials, we make a digression here
and explain the relation between magnetizations
as evaluated using the physical electron-picture (as in Fig.~\ref{hybrid}) 
and magnetizations evaluated using the indirect but computationally more convenient hole-picture.
  
\begin{figure}[h]
\includegraphics[width=2.8in,angle=0]{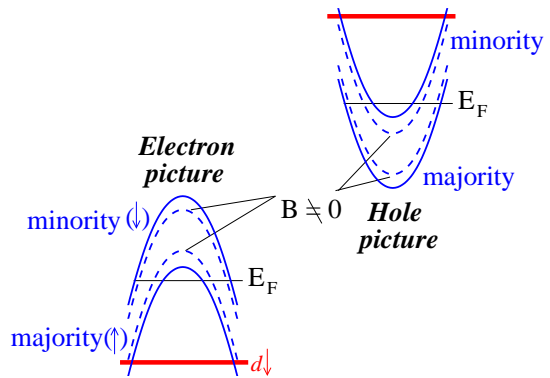}
\caption{Cartoon of Zeeman coupling of an external magnetic field assuming $g>0$ in the 
electron and hole pictures for our valence band coupled to Mn moments system.
Majority band in both electron and hole pictures moves up in energy resulting
in a negative band-contribution to the magnetization.} 
\label{el-hole-pict}
\end{figure}

Magnetization at $T=0$ is defined thermodynamically by the 
dependence of the ground-state energy
$E$ on external magnetic field $B$:
\begin{equation}
m=\left. -\frac{\partial E}{\partial B}\right|_{B=0}
\label{magdef}
\end{equation}
In this paper we always assume ${\bf B}\parallel +\,\hat{\bf z}$.
In a mean-field picture the magnetization is related to the
change of single-particle energy with field, summed over all occupied orbitals.
Orbitals that decrease in energy with field make a positive contribution to the 
magnetization.  For ${\bf B}\parallel +\,\hat{\bf z}$, the $d$-electron spins are
aligned along $-z$-direction (down-spins) and the majority spin band electrons have spin-up due
to antiferromagnetic $p-d$ exchange coupling. Then, if the majority spin band 
moves up in energy with $B$ and the minority band moves down, as illustrated
in the left part of Fig.~\ref{el-hole-pict}, the band kinetic energy increases with $B$ and, according to
Eq.~(\ref{magdef}), the corresponding contribution to the magnetization is negative.
In the hole-picture, we obtain 
the same respective sense of shifts of the majority {\em hole} and minority {\em hole}  
bands, as illustrated in the right part of Fig.~\ref{el-hole-pict}, and therefore the correct (negative in our case) sign
of the magnetization. The cartoon shows that in order to 
circumvent the rather confusing notion of a spin of a hole in magnetization calculations, the full Hamiltonian $\hat{H}(B)$
should be derived in the physical electron picture where the sign of the coupling of the electron spin to $B$
is unambiguously defined. The electron picture $\rightarrow$ hole picture transformation ($\hat{H}(B)\rightarrow
-\hat{H}(B)$)
and the clearly defined notion of majority 
and minority bands in either picture guarantees the consistency in sign of the calculated magnetization.
Note that the language used here neglects spin-orbit interactions which lead to single-particle
orbitals that do not have definite spin character.  Although spin-orbit interactions are important
they can be neglected in most qualitative considerations, like the ones we explain here.  
In this paper we occasionally make statements which neglect spin-orbit interactions, and they
should always be understood in this spirit. 

The electron-electron exchange energy has a negative sign and its magnitude increases monotonically when
moving from the paramagnetic
to the half-metallic (empty minority band) state. This together with Eq.~(\ref{magdef}) implies
that the magnetization contribution from the electron-electron
exchange energy has the same sign as the contribution from the kinetic energy. Using the same arguments as above
we see in the exchange energy case the sign of magnetization is also treated consistently in 
the electron picture $\rightarrow$  hole picture transformation.

\subsection{Mean-field magnetization}
In a model which for the many-Mn-impurity system corresponds to a mean-field approximation,
the ground state wavefunction of the Mn($d^5$+hole) complex reads $|S_z=-S\rangle|j_z=+j\rangle$ and
the magnetization per Mn equals $m_{MF}=(g_SS-g_jj)\mu_B$, 
where $S$ and $j$ are local $d$-electrons and hole
moments and $g_S$ and $g_j$ are the respective Land\'e g-factors. The five $d$-electrons have
zero total orbital angular momentum, i.e. $g_S=2$, and for spin $j=1/2$ hole ($g_j=2$) we get $m_{MF}=4\mu_B$. 
Hole states near the valence band edge have a $p$-character, however, so more realistically
we should consider $g_j j=4/3 * 3/2=2$ which
gives $m_{MF}=3\mu_B$. We show in the next section that this basic picture of a 
suppressed $m_{MF}$ 
due to holes applies also to highly Mn-doped (Ga,Mn)As materials although the magnitude of the 
mean-field hole
contribution is weaker because
of the occupation of both majority and minority hole bands and, partly, because of spin-orbit coupling effects.   

\subsection{Quantum fluctuations contribution to magnetization}

The two-spin, $S$ and $j$ model allows us also  to  readily demonstrate the presence of
quantum fluctuations around the mean-field ground state, which is related to the 
antiferromagnetic sign of the ${\bf S}\cdot {\bf j}$ coupling. We show that for the Mn($d^5$+hole) complex,
quantum fluctuations are expected
to weakly suppress the mean-field magnetization $m_{MF}$. Detailed many-body calculations discussed in the 
following section confirm the role of quantum fluctuations is also weak for the many-Mn systems.

In the limit of $B\rightarrow 0$ we can write  the two-spin Hamiltonian as,
\begin{equation}
{\mathcal H}=J \;{\hat {\bf S}}\cdot {\hat {\bf j}}=\frac{J}{2} \, (\hat{S}_{tot}^2-\hat{S}^2-\hat{j}^2)\, ,
\label{two-spin-Ham}
\end{equation}
where ${\hat {\bf S}}_{tot}={\hat {\bf S}}+{\hat {\bf j}}$. For comparison we first consider  
ferromagnetic coupling, $J<0$. Since in this case $S_{tot}=S+j$,  
the ground-state eigenenergy, $E_{FM}=-\frac{|J|}{2}[(S+s)(S+j+1)-
S(S+1)-j(j+1)]=-|J|Sj$, equals to the mean-field energy, i.e., the mean-field state
is exact.  For
ferromagnetic ${\bf S}\cdot {\bf j}$ coupling, the quantum fluctuation contribution
to the magnetization is strictly absent in the many-Mn case 
only when the hole system is half-metallic ({\em i.e.} when the minority band is empty).
We can see this by introducing spin raising and lowering operators
in  the Hamiltonian~(\ref{two-spin-Ham}),
\begin{equation}
{\mathcal H}=J\left[\hat{S}_{z}\hat{j}_z+\frac12(\hat{S}^+\hat{j}^-+\hat{S}^-\hat{j}^+)\right]\;.
\label{two-spin-Ham_2}
\end{equation}
Quantum fluctuations are absent when the transverse spin terms above annihilate the
many-particle ground state.  When acting on a state with all localized spins {\em and} all band spins polarized 
in the same direction, both transverse terms produce zero.  For partially spin-polarized bands,
quantum fluctuation corrections, although not strictly zero, are 
qualitatively smaller than in the antiferromagnetic case.  

For antiferromagnetic coupling ($J>0$), $S_{tot}=S-j$ and
the corresponding ground-state energy $E_{AF}=\frac{|J|}{2}[(S-j)(S-j+1)-
S(S+1)-j(j+1)]=-|J|(Sj+j)$ is lower than the mean-field energy.
The mean-field ground
state is not exact here and quantum fluctuation corrections to the
magnetization will be non-zero in general. To estimate the correction we write
the exact ground-state
wavefunction as,
\begin{eqnarray}
|\psi\rangle&=&|S_{tot}=S-j,S_{tot,z}=-(S-j)\rangle\nonumber\\
&=&\sqrt{\frac{S}{S+j}}|S_z=-S,j_z=+j\rangle \nonumber\\
& &
-\sqrt{\frac{j}{S+j}}|S_z=-S+1,j_z=+j-1\rangle\; ,
\label{exact_AF}
\end{eqnarray}
the mean-field wavefunction as,
$|\psi\rangle_{MF}=|S_z=-S,j_z=+j\rangle$, and evaluate the respective expectation
values of the Zeeman Hamiltonian, $g_S\mu_BB\hat{S}_z+g_j\mu_BB\hat{j}_z$. From Eq.~(\ref{magdef})
we then obtain that the difference between the exact and mean-field state magnetizations
is given by
\begin{equation}
m-m_{MF}\equiv m_{QF}=-\mu_B
\frac{j}{S+j}(g_S-g_j)
\; .
\label{qf_correction}
\end{equation}
When $j=1/2$ and $g_S=g_j=2$ the quantum fluctuation correction to the magnetization vanishes
even though the mean-field ground state is not exact.  The quantum fluctuation correction
to the magnetization remains relatively weak also even when we adopt the more realistic description of the valence-band hole
moment ($j=3/2$, $g_j=4/3$), for which 
$m_{QF}=-0.25\mu_B$. 

\section{Magnetization of many-${\rm\bf Mn}$-impurity system}
\label{manyMn}
\subsection{Moments per Mn in the microscopic TBA/CPA model}
\label{AF-TBA}
The TBA description of (Ga,Mn)As mixed crystals is particularly useful 
for explaining the complementary role of local and
itinerant moments in this p-type magnetic semiconductor.
The language that is used to describe this interplay can differ depending 
on whether a fully microscopic or a kinetic-exchange model is employed, 
and this difference has sometimes led to confusion. 
This section represents an attempt at clarity.  At the same time
we find that the TBA/CPA results help establish the  validity of the antiferromagnetic $p-d$ kinetic-exchange model and of the
virtual crystal approximation for describing collinear ground states in highly
doped (Ga,Mn)As ferromagnets.\cite{Jungwirth:2003_c,Jungwirth:2005_b} 

\begin{figure}[h]
\includegraphics[width=3.6in,angle=0]{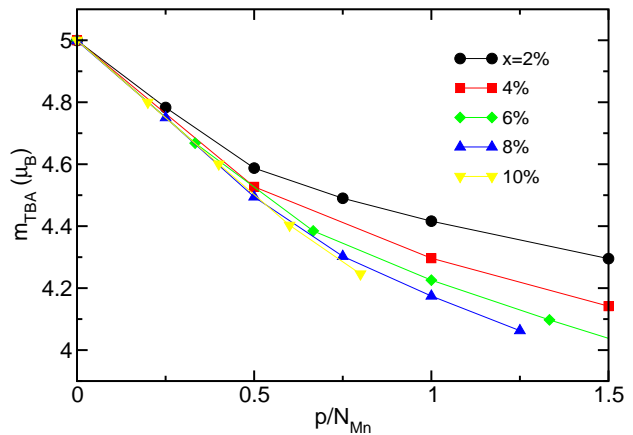}
\caption{Mean-field total magnetization per Mn as a function of  hole density relative to the local
Mn moment density. Results are obtained using the TBA/CPA model.
} 
\label{tba}
\end{figure}

In the TBA/CPA calculations, the hole density is varied independently of 
Mn doping by adding non-magnetic donors (Si or Se) or acceptors (C or Be).
The $d$-electron magnetic moments of all Mn atoms are aligned
along  $+z$-axis. The parameterization  
of the TBA Hamiltonian 
was chosen to provide  the correct band gap
for  a  pure  GaAs  crystal \cite{Talwar:1982_a}  and  the  appropriate
exchange splitting of  the Mn $d$-states. Local changes  of the crystal
potential at Mn, represented
by     shifted     atomic      levels,     were    estimated     using
Ref.~\onlinecite{Harrison:1980_a}. Spin-orbit coupling is not included
in our TBA Hamiltonian. 
In the CPA, disorder effects  appear in  the finite  spectral width  of  hole
quasiparticle states.  The TBA/CPA technique can, therefore, capture
changes in the $p-d$ 
interaction with doping due to both chemical alloying effects and positional disorder.
In Fig.~\ref{tba} we show the microscopic TBA/CPA magnetic moments per Mn, $m_{TBA}$, in
Ga$_{1-x}$Mn$_x$As ferromagnets plotted as a function of the hole density $p$
relative to the Mn concentration $N_{Mn}=4x/a_{\rm lc}^3$ ($a_{\rm lc}$
is the semiconductor host lattice constant). The $m_{TBA}$ is obtained here using the electron picture by
integrating over occupied states up to the Fermi energy. 

A common way of microscopically separating 
contributions from local atomic and itinerant moments is by projecting the occupied electron states
onto Mn $d$-orbitals and $sp$-orbitals, respectively. In this decomposition, the resulting local Mn moments are 
smaller than 5$\mu_B$ per Mn  due to the admixture of $d$-character in empty states near the valence band edge. 
The effective kinetic-exchange model employed in the following sections corresponds, however, to a
different decomposition of contributions, in effect associating one spectral region with local Mn moments and 
a different spectral region with itinerant hole moments.
The kinetic-exchange model, in which local moments have $S=5/2$, is obtained from the 
microscopic TBA/CPA model by expressing the total TBA/CPA moment as the difference between a contribution $m^{int}_{TBA}$ resulting from
integrating over all electronic states up to mid-gap, {\em i.e.} including the entire valence band, and
a contribution corresponding to the integral from Fermi energy to mid-gap. As long as the valence-conduction
band gap is non-zero, the former contribution is independent of valence band filling and equals to the moment of an isolated Mn atom, 5$\mu_B$.
The latter term represents magnetization of itinerant holes.

\begin{figure}[h]
\includegraphics[width=3.6in,angle=0]{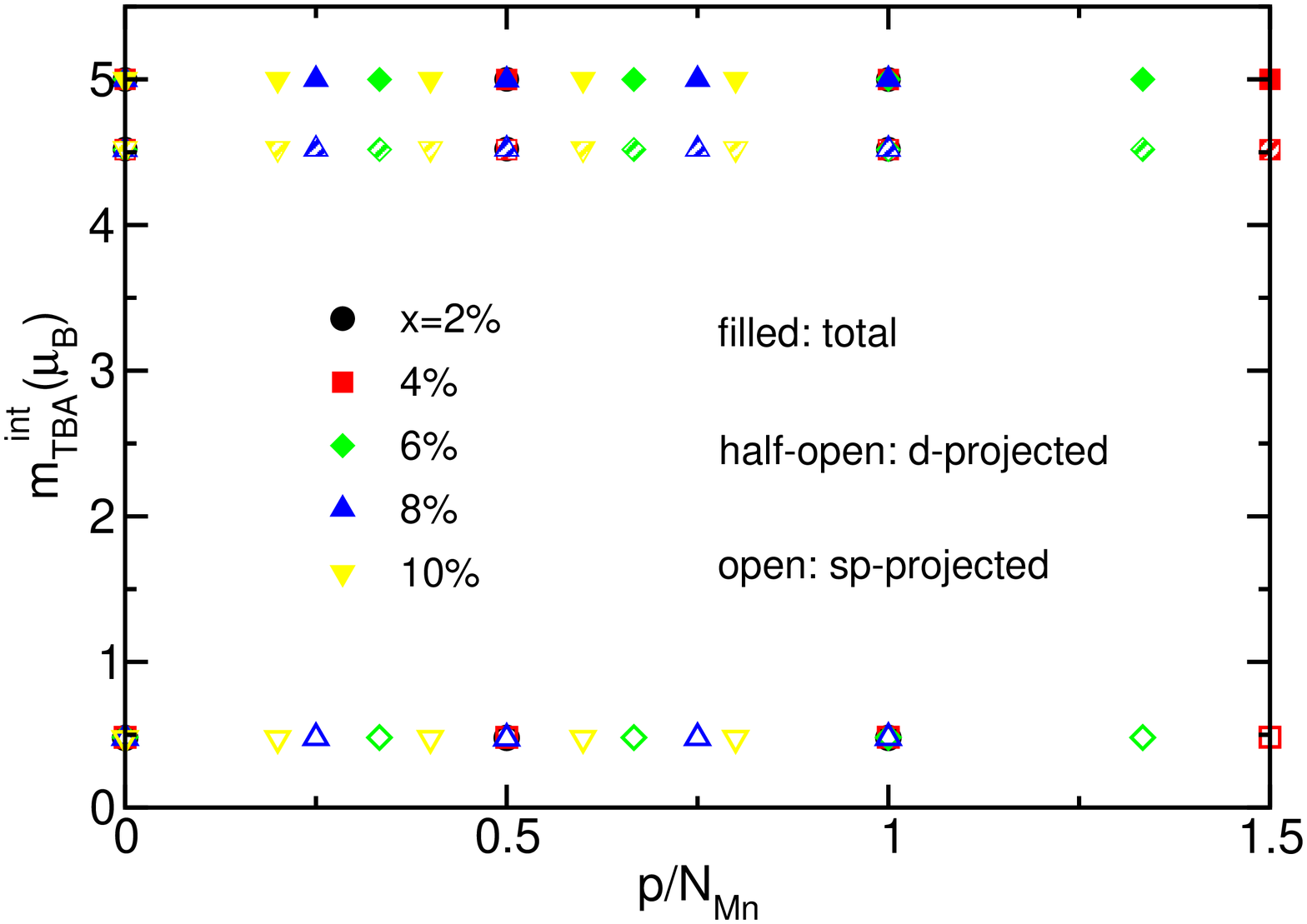}
\caption{Integrated total and $d$- and $pd$-projected magnetizations 
per Mn as a function of  hole density relative to the local
Mn moment density. See text for definition of $m^{int}_{TBA}$.
} 
\label{tba2}
\end{figure}

The applicability of the effective kinetic-exchange model relies implicitly on the 
perturbative character of the microscopic
$p-d$ hybridization. The level of the $p-d$ hybridization over the studied doping range
is illustrated in Fig.~\ref{tba2} where we show the
orbital composition of $m^{int}_{TBA}$. The filled symbols correspond
to including spectral weights from all $spd$ orbitals while the
half-open and open symbols are obtained after projecting onto the
$d$ and $sp$ orbitals, respectively.  If no hybridization was
present, then  $m^{int}_{TBA}$ projected on the $d$-orbitals would
equal to the total $m^{int}_{TBA}$ and the $sp$-orbital projected
$m^{int}_{TBA}$ would vanish. In our TBA/CPA calculations, the
$d$-orbital projected  $m^{int}_{TBA}$ is reduced by only 10\% as
compared to the total $m^{int}_{TBA}$ and, therefore, the $p-d$
hybridization can be regarded as a weak perturbation. The nearly constant
value of the $d$-orbital projected $m^{int}_{TBA}$ also suggests
that the kinetic-exchange coupling parameter $J_{pd}$ in the effective spin
Hamiltonian is nearly independent of doping over the whole range of
Mn and hole densities that we study.

The decrease of
$m_{TBA}$ in Fig.~\ref{tba} with increasing $p/N_{Mn}$ clearly demonstrates
antiferromagnetic $p-d$ coupling over the whole range
of dopings. The initial common slope for data corresponding to different 
Mn concentrations reflects the half-metallic nature
of the hole system (only majority hole band occupied) when spin-orbit interactions
are neglected. Here the hole contribution
to magnetization per volume is proportional to $p$, i.e., magnetization per Mn is
proportional to $p/N_{Mn}$. The change in the slope of $m_{TBA}$ at larger hole densities, which now becomes
Mn-density dependent,  
reflects population of the minority-spin hole band and, therefore, the additional dependence of hole magnetization
on exchange splitting between majority- and minority-hole bands. 
In this regime the hole magnetization per volume is approximately
proportional to $pN_{Mn}$, i.e., a common trend for different Mn densities is obtained when $m_{TBA}$
is plotted versus $p$ rather than $p/N_{Mn}$. 

Within the TBA/CPA model, $m_{TBA}$ corresponds
to the  total magnetization per Mn measured by a SQUID and  
we therefore expect these experimental values to lie between 4 and 5
$\mu_B$ and to increase with decreasing number of holes in collinear (Ga,Mn)As ferromagnets. 
The XMCD data on the other hand reflect
local and d-state projected contribution from Mn and should be compared with the half-open symbols
in Fig.~\ref{tba2}, showing a negligible dependence on the hole density. These trends are indeed confirmed in our 
experiments. Before discussing the experimental data we refine, quantitatively, the above theoretical predictions for the
total magnetization. For example, we expect that the number of minority holes at a given total hole density
is underestimated in our TBA model. This is caused in part by
the quantitative value of the exchange spin-splitting of the valence band in the TBA/CPA calculations 
which is a factor of 1.5-2 larger than 
value inferred from experiment. Also since SO coupling is not included in
our TBA model, we obtain three majority bands that are degenerate
at the $\Gamma$-point, instead of only two bands (heavy- and light-hole) as in the
more realistic SO-coupled band structure.  (This substantial deficiency is common
to all calculations that neglect spin-orbit coupling.) In addition to the underestimate
of the minority hole density, the TBA/CPA calculations also omit the reduction
of the mean spin-density
in the majority band caused by the SO-coupling and, therefore, the total TBA 
magnetization values  are too small.
In the following section we attempt to correct for 
these quantitative shortcomings of the TBA calculations by
taking the experimentally 
measured\cite{Okabayashi:1998_a,Omiya:2000_a} value of the $p-d$ coupling constant,
$J_{pd}=54\pm 9$~meV~nm$^3$, and by 
evaluating valence band spin-splitting  and SO coupling effects within 
the semiphenomenological ${\protect\rm k}\cdot {\protect\rm p}$ 
kinetic-exchange model.\cite{Dietl:2001_b,Abolfath:2001_a} The weak dependence
of the TBA/CPA valence band splitting on positional disorder justifies our use of  
the virtual crystal approximation in this semiphenomenological modeling of collinear
(Ga,Mn)As ground states. 

\subsection{Mean-field magnetization contributions from hole kinetic and exchange energies}
\label{hole-mag}
Within the semiphenomenological virtual crystal model the valence band holes
experience a mean-field, $h_{MF}   =  J_{pd}   N_{Mn} \langle S\rangle$, and the band Hamiltonian can then
be written as, $\hat{H}_{MF}=\hat{H}_{KL}(B)+ h_{MF}\hat{s}_z$,
where  $\hat{H}_{KL}(B)$ is the $B$-dependent six-band Kohn-Luttinger Hamiltonian of  the  GaAs host
band\cite{Dietl:2001_b,Vurgaftman:2001_a}, $\hat{s}_z$   is   the  $z$-component of
the hole spin
operator, and $\langle S\rangle$ is the mean spin polarization of the local 
Mn moments.\cite{Dietl:2000_a,Dietl:2001_b,Abolfath:2001_a} 
At $T=0$, $\langle S\rangle=5/2$ and the local moment contribution to the magnetization
per Mn is 5$\mu_B$. 
As emphasized above, this finding is {\em not} in contradiction with the smaller d-electron
contribution to the magnetic moment in microscopic calculations.  

\begin{figure}[h]
\includegraphics[width=3.6in,angle=0]{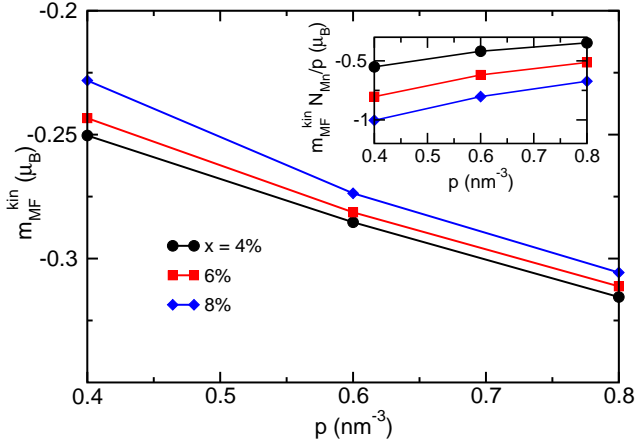}
\caption{Mean-field kinetic energy contribution to the hole magnetization per Mn as a function of hole density.
These results were obtained using the six-band Kohn-Luttinger parameterization of the valence band
and the kinetic-exchange model.
} 
\label{kin_mf}
\end{figure}

Because of the SO interaction, both orbital and Zeeman
couplings of the external magnetic field have to be included in $\hat{H}_{KL}(B)$. The SO-coupling
and heavy-hole -- light-hole mixing at finite wavevectors lead to magnetizations that cannot be expressed
using a constant, Mn- and hole-density independent g-factor. Instead the kinetic
band energy contribution to MF magnetization per Mn,
$m^{kin}_{MF}$, is obtained by numerically integrating 
over all occupied hole eigenstates of $\hat{H}_{MF}$ and by finding the 
coefficient linear in $B$ of this kinetic energy contribution to the total energy.\cite{Dietl:2001_b} 
Results for several typical Mn dopings 
and hole densities are shown in Fig.~\ref{kin_mf}. They agree quantitatively with earlier calculations reported
in Ref.~\onlinecite{Dietl:2001_b}. As anticipated in Section~\ref{d5+hole},
$m^{kin}_{MF}$ is negative, {\em i.e.} it suppresses the total magnetic moment. The 
magnitude of the hole MF magnetization per hole,
$|m^{kin}_{MF}|N_{Mn}/p$, is smaller than 2$\mu_B$ due to occupation of both majority and minority heavy- and
light-hole bands at these typical (Ga,Mn)As hole densities (see inset of
Fig.~\ref{kin_mf}). In this case, as also emphasized in  Section~\ref{d5+hole},
$m^{kin}_{MF}$ is expected to fall into a common trend for different Mn densities when plotted against $p$.
Data shown in the main panel of Fig.~\ref{kin_mf} confirm this expectation and indicate
a ~$\sim$0.2 to 0.4$\mu_B$ suppression of the MF moment per Mn due to the hole kinetic
energy contribution to magnetization.\cite{Dietl:2001_b} 

To estimate the hole exchange energy contribution to magnetization per Mn, $m^{ex}_{MF}$, 
we use an expression of the
total exchange energy derived in the absence of SO-coupling and assuming spin-up and spin-down
heavy-hole bands with effective mass 0.5$m_e$,
 \begin{equation}
E_{ex}=2^{1/3}\frac{E^P_{ex}(n)}{p^{4/3}}\left[
p_{\uparrow}^{4/3}(B)+
p_{\downarrow}^{4/3}(B)\right]\,,
\label{exchange_en}
\end{equation}
where $p_{\uparrow(\downarrow)}$ is the density of the majority(minority)
band, $p_{\uparrow}+p_{\downarrow}=p$, and the exchange energy of the paramagnetic state is given by
\begin{equation}
E^P_{ex}(n)=-\frac{e^2}{4\pi\varepsilon}\frac34
\left(\frac{3}{\pi}\right)^{1/3}p^{4/3}\; .
\end{equation}
The $B$-dependent hole densities can be written as
\begin{eqnarray}
p_{\uparrow}(B)=p_{\uparrow}(0)-|g_j\mu_B B j|
\frac{2G^{\uparrow}G^{\downarrow}}{G^{\uparrow}+G^{\downarrow}}\nonumber \\
p_{\downarrow}(B)=p_{\downarrow}(0)+|g_j\mu_B B j|
\frac{2G^{\uparrow}G^{\downarrow}}{G^{\uparrow}+G^{\downarrow}}\; ,
\label{density}
\end{eqnarray}
where $G^{\uparrow(\downarrow)}$ is the density of states at the Fermi energy of the
majority(minority) band. Combining Eqs.~(\ref{exchange_en})-(\ref{density}) and Eq.~(\ref{magdef})
we obtain
\begin{equation}
m^{ex}_{MF}=
\frac43\frac{2^{1/3}}{p^{4/3}}E^P_{ex}(n)
g_j\mu_Bj\frac{2G^{\uparrow}G^{\downarrow}}{G^{\uparrow}+G^{\downarrow}}
\left(p_{\uparrow}(0)^{1/3}-p_{\downarrow}(0)^{1/3}\right)\, .
\label{mex}
\end{equation}

\begin{figure}[h]
\includegraphics[width=3.6in,angle=0]{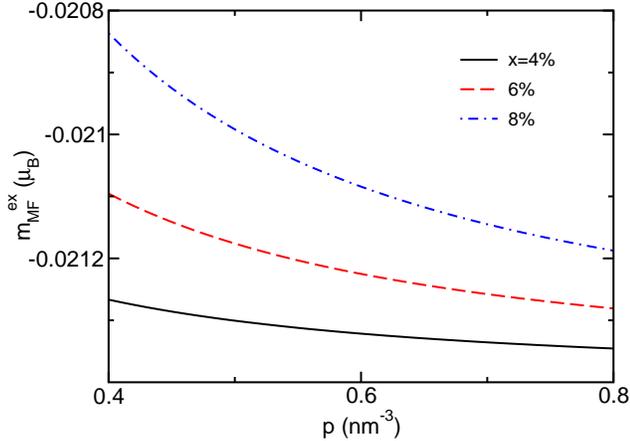}
\caption{Mean-field hole-hole exchange energy contribution to the hole magnetization per Mn as a function of hole density.
These results were obtained using the spin-1/2, $m^{\ast}=0.5m_e$ parabolic band kinetic-exchange model.
} 
\label{exch_mf}
\end{figure}

\begin{figure}[h]
\includegraphics[width=3.6in,angle=0]{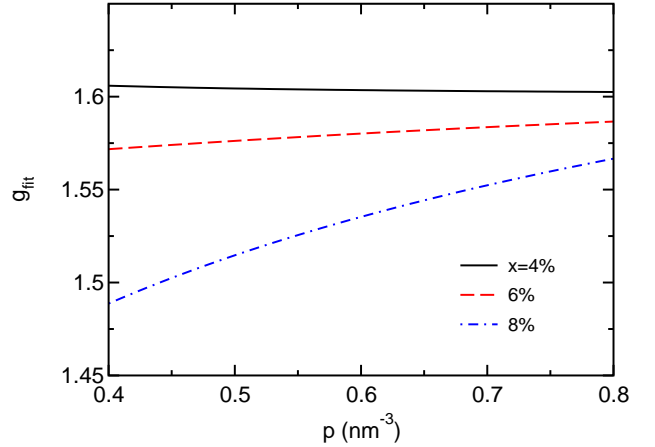}
\caption{Effective hole Land\'e g-factor obtained by fitting the kinetic energy term in hole magnetization calculated 
with the parabolic band kinetic-exchange model with $j=3/2$ and $m^{\ast}=0.5m_e$ to the numerical results of the six-band Kohn-Luttinger
model from Fig.~\protect\ref{kin_mf}} 
\label{gsfit}
\end{figure}

In Fig.~\ref{exch_mf} we plot $m^{ex}_{MF}$ values calculated assuming spin-1/2 holes, {\em i.e.} $g_jj=1$. 
This contribution to the
total MF magnetization is again negative, is nearly independent of $x$ and $p$ within the
range of doping considered, and its magnitude is about a factor of 10 smaller than the magnitude of
the term originating from the hole kinetic band energy. 
A more realistic estimate of $m^{ex}_{MF}$ can be obtained by using
$g_jj=g_{fit}*3/2$ in Eq.~(\ref{mex}). Here $g_{fit}$ follows from fitting the $m^{kin}_{MF}$ calculated within the
parabolic heavy-hole band model to  the full six-band numerical results in Fig.~\ref{kin_mf}. The values of $g_{fit}$ as a function
of $x$ and $p$ are plotted in Fig.~\ref{gsfit}. These are similar to the $g_{j=3/2}=4/3$ value that follows from the local
atomic model and, therefore, 
$m^{ex}_{MF}$ calculated from Eq.~(\ref{mex}) using $g_jj=g_{fit}*3/2$ is approximately a factor of 2 larger than
$m^{ex}_{MF}$ calculated assuming spin-1/2 holes.  Combining all these considerations we conclude
that the zero-temperature magnetization per Mn in the
MF kinetic-exchange model has a positive contribution equal to 5$\mu_B$ from the Mn local moments and a negative 
contribution from band holes which suppresses the moment per Mn by $\sim$5-10\%.

\subsection{Quantum fluctuation contribution to the magnetization}
\label{qf-mag}
In Section~\ref{d5+hole} we argued that quantum fluctuation corrections to the 
isolated Mn $d^{5}$+hole complex magnetization should be small. Here we demonstrate that this
conclusion also applies to the many-Mn system.
In these calculations we use the virtual crystal kinetic-exchange model and assume spin-1/2 heavy-holes with no
SO-coupling and with the parabolic band dispersion ($m^{\ast}=0.5m_e$).
The many-body Hamiltonian of the model reads
\begin{eqnarray}
  \hat H &=& \int d^3 r \left[ \sum_\sigma
      \hat\Psi^\dagger_\sigma (\vec r) \left( -{\hbar^2 \vec\nabla^2\over 2m^*}
        -\mu \right) \hat \Psi_\sigma (\vec r)
        + g_j \mu_B \vec B \cdot \vec j (\vec r) \right.
\nonumber  \\ 
        &+& \left. |J_{pd}| \sum_I \vec S(\vec R_I) \cdot \vec j(\vec r)
        \delta (\vec r - \vec R_I) \right] \nonumber \\
        &+& g_S\mu_B \sum_I \vec B \cdot \vec S(\vec R_I)\;.
\label{many_body_H}
\end{eqnarray}

The imaginary time path-integral formulation of quantum statistical physics combined
with a Holstein-Primakoff bosonic representation
for the Mn local moments allows\cite{Konig:2001_a} us to 
formally express the free energy of interacting local and itinerant spins in terms of a path 
integral over coherent state labels ${\bar z},z$: 
\begin{equation}
{\mathcal Z}=\int {\mathcal D}({\bar z}z)\exp{\left(-S_{\rm eff}[{\bar z}z]\right)}\,.
\label{part_function}
\end{equation} 
The effective action $S_{\rm eff}$ in 
Eq.~(\ref{part_function}) is obtained by integrating out fermionic (hole) degrees
of freedom in Eq.~(\ref{many_body_H}).  In the Gaussian fluctuation approximation\cite{Konig:2001_a}
\begin{equation}\label{quadratic}
S_{\rm eff}[{\bar z}z]=\frac{1}{\beta}\sum_m\int_{|{\bf q}|\leq q_c} 
\frac{d^3{\bf q}}{(2 \pi)^3}\,{\bar z}({\bf q},\nu_m)D^{-1}({\bf q}, i\nu_m)z({\bf q},\nu_m)\; .
\end{equation}
Here the inverse of the spin-wave propagator $D({\bf q}, \nu_m)$ is given by
\begin{equation}\label{swp}
D^{-1}({\bf q}, i\nu_m)=-i \nu_m +\varepsilon_{sw}(B)+\Sigma_{sw}({\bf q}, i\nu_m,B)\,,
\end{equation}
$q_c=(6\pi^2 N_{Mn})^{1/3}$ is a Debye cutoff which 
ensures that we include the correct number of local-moment degrees of freedom and
\begin{equation}\label{epsilon}
\varepsilon_{sw}(B)=-g_S\mu_B B+\frac{|J_{pd}|}{2}\left(p_{\uparrow}(B)-p_{\downarrow}(B)\right)\,
\end{equation}
is the mean-field local moment spin-flip energy.  
The frequency-dependent self-energy $\Sigma_{sw}({\bf q}, i\Omega,B)$ in Eq.~(\ref{swp})
is given by
\begin{eqnarray}\label{self_energy}
& &\Sigma_{sw}({\bf q}, i\Omega,B)=\frac{N_{Mn} J_{pd}^2 S}{2}\nonumber \\
& &\times\int \frac{d^3 {\bf k}}{(2\pi)^3}
\frac{f(\varepsilon_{\bf k}-\mu +\Delta(B)/2) - 
f(\varepsilon_{{\bf k} + {\bf q}}-\mu-\Delta(B)/2)}{i\Omega+\varepsilon_{\bf k}-\varepsilon_{{\bf k}+{\bf q}}+\Delta(B)}\nonumber\,,
\end{eqnarray}
with $\varepsilon_{\bf k}=\hbar^2k^2/2m^*$ and 
\begin{equation}\label{delta}
\Delta(B)=N_{Mn}|J_{pd}|S-g_j \mu_B B\,.
\end{equation}
The translational and rotational invariance of our model implies that $\Sigma_{sw}({\bf q}, i\Omega)$ depends only of $|{\bf q}|$. 
The functional integration in Eq. (\ref{part_function}) can be performed 
exactly using Eq.~(\ref{quadratic}),
\begin{equation}
{\mathcal Z}=\frac{1}{\beta}\prod_{n, |{\bf q}|\leq q_c}\frac{1}{-i\nu_n+
\varepsilon_{sw}(B)+\Sigma_{sw}(q, i\nu_n,B)}\,.
\label{Z-final}
\end{equation}
The quantum fluctuation 
correction to the free energy then reads,
\begin{equation}
\delta F_{\rm QF}\equiv F-F_0=-\frac{1}{\beta}\ln{\frac{{\mathcal Z}}{{\mathcal Z}_0}}\,,
\end{equation} 
where ${\mathcal Z}_0$ is given by Eq.~(\ref{Z-final}) with the fluctuations term, 
$\Sigma_{sw}(q, i\nu_n,B)$, set to zero. 

\begin{figure}[h]
\includegraphics[width=3.2in,angle=0]{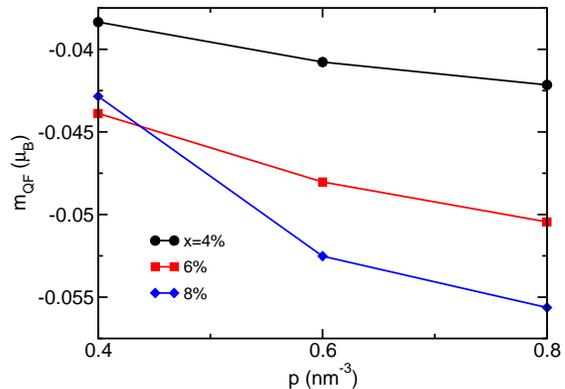}
\caption{Quantum fluctuation contribution to  magnetization per Mn as a function of hole density.
These results were obtained using the spin-1/2,  $m^{\ast}=0.5m_e$ parabolic band kinetic-exchange model.} 
\label{fluct}
\end{figure}

In Fig.~\ref{fluct} we plot the quantum fluctuations contribution to the magnetization per Mn obtained
from 
\begin{equation}
m_{QF}=-\left.\frac{\partial \delta E_{\rm QF}}{\partial B}\right|_{B=0}\,.
\end{equation} 
As expected $m_{QF}$ is small in the many-Mn system and we can conclude that quantum fluctuations lead
to a $\sim$1\% suppression of the MF moment per Mn.

\section{Experimental data}
\label{experiment}
\subsection{Sample growth and preparation}
In the theoretical sections of this paper we evaluated the zero-temperature 
magnetization per Mn in (Ga,Mn)As ferromagnets, using approximation schemes
that would fail if the true ground state magnetization was highly non-collinear. We will now 
show that these theoretical results are consistent with low temperature magnetometry and XMCD experiments. 
A series of (Ga,Mn)As films with Mn content varying between 1.7-6.7\% in the SQUID experiments
and between 2.2 and 8.4\% in the XMCD experiments
were grown by low-temperature molecular beam epitaxy (MBE)
using As$_2$. The layer structure of the thin films consists of 
25 or 50nm (Ga,Mn)As / 50nm low temperature GaAs / 100nm high temperature (580$^{\circ}$C) GaAs / SI-GaAs(100) substrate.
The growth temperature of the (Ga,Mn)As layer and the GaAs buffer was 180-300$^{\circ}$C, decreasing with increasing Mn concentration in order to minimize As antisite densities while maintaining two-dimensional growth and preventing phase segregation. Further details on the growth are presented elsewhere.\cite{Campion:2003_a,Foxon:2004_a} 

The Mn concentrations were deduced from the in-situ measured Mn/Ga incident flux ratio, which was calibrated using secondary ion mass spectrometry (SIMS) measurements on 1$\mu$m thick (Ga,Mn)As films, grown under otherwise identical conditions to the samples considered here. A detailed comparison of the results of a number of different calibration techniques, presented in detail elsewhere,\cite{Zhao:2005_b} allows us to assign an uncertainty of ±10\% to the quoted total Mn doping, $x$. The SIMS measurements yield no information on the lattice site of the incorporated Mn, and we expect that
Mn will be incorporated either on interstitial Mn$_{\rm I}$ or on 
substitutional Mn$_{\rm Ga}$ sites.\cite{Yu:2002_a} 
Post-growth annealing of the samples is performed in air at 190$^{\circ}$C for 50-150 hours, which is an established procedure for removal of  Mn$_{\rm I}$ from the (Ga,Mn)As layer. Curie temperatures in the as-grown materials are 
within the range of 40-80~K and in the annealed samples between 40 and 150~K.\cite{Jungwirth:2005_b}

\subsection{Magnetometry}
\begin{figure}[h]
\hspace*{-1cm}\includegraphics[width=3.0in,angle=-90]{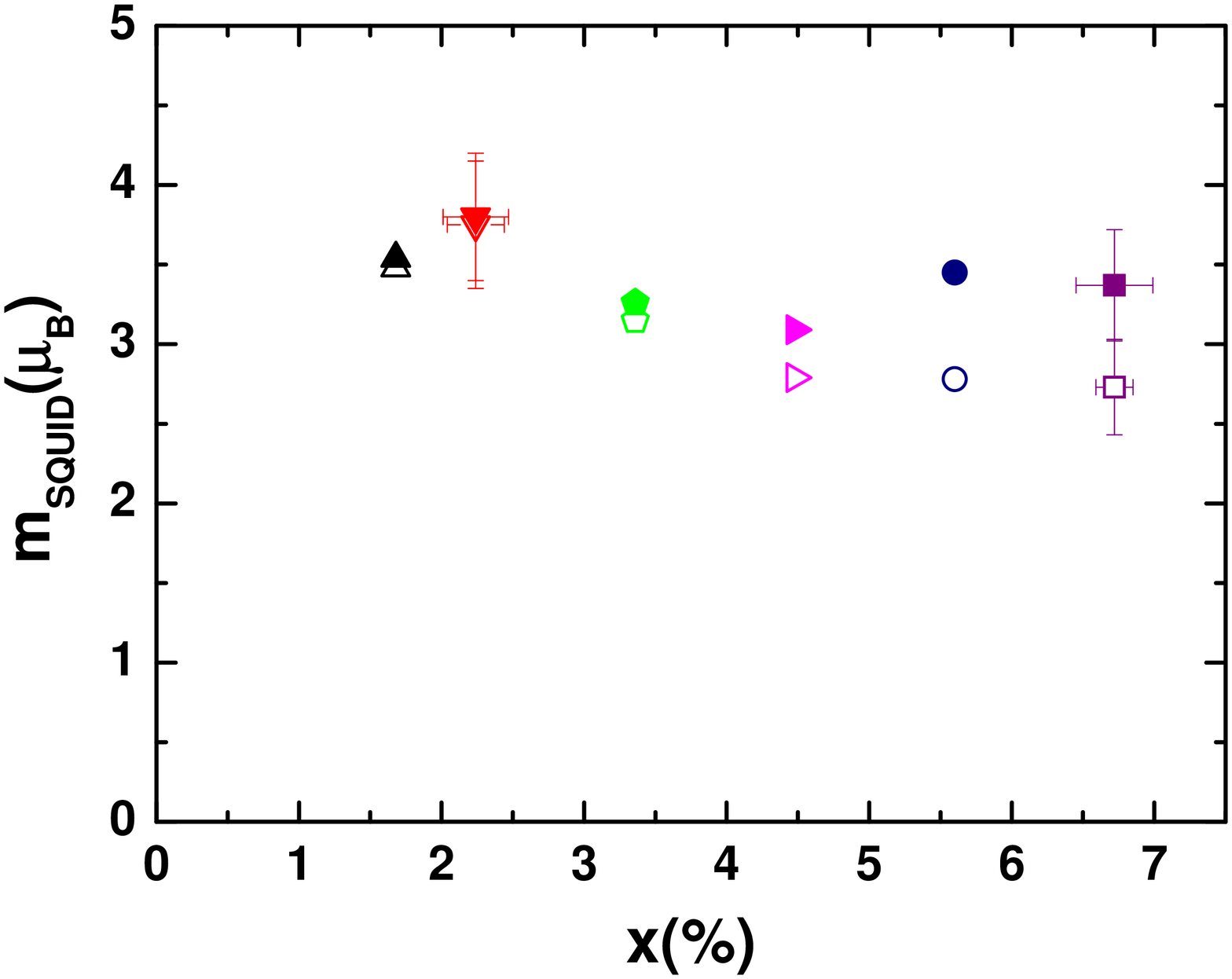}
\caption{SQUID magnetization per nominal total Mn density in as-grown (open symbols) and
annealed (filled symbols) (Ga,Mn)As materials plotted as a function of the nominal Mn doping.} 
\label{M-x}
\end{figure}

The magnetic moment of the samples is measured in a SQUID magnetometer, at 5K and under a 0.3T external magnetic field. The external  field is necessary to overcome  in-plane anisotropy fields, so that the magnetization 
is aligned with the measurement axis of the SQUID. The diamagnetic contribution from the substrate is subtracted.
Measured magnetic moments normalized to the total Mn concentration as obtained from SIMS calibration, 
$m_{SQUID}$, are shown in Fig.~\ref{M-x}. The moment decreases with increasing Mn concentration, and increases on annealing, similar to earlier reports.\cite{Potashnik:2002_a} This is consistent with the anticipated formation of interstitial Mn for doping above $\sim$2\%,\cite{Jungwirth:2005_b} 
given the  antiferromagnetic coupling between  Mn$_{\rm I}$ 
and Mn$_{\rm Ga}$,\cite{Blinowski:2003_a} and with breaking of this coupling by 
low-temperature annealing.\cite{Yu:2002_a,Edmonds:2004_a}

In order to compare the experimental data with the theoretical results of previous sections we have to
replot the measured magnetizations as a function of the density of uncompensated Mn$_{\rm Ga}$ local moments, 
$x_{eff}$.  To do this we need  to determine the densities of substitutional Mn$_{\rm Ga}$ and interstitial Mn$_{\rm I}$,
$x_s$ and $x_i$, in our (Ga,Mn)As materials. 
Given these values, we assume\cite{Jungwirth:2005_b} that each Mn$_{\rm I}$ present in the system
is antiferromagnetically coupled to one Mn$_{\rm Ga}$ and that both should 
be excluded from the active Mn fraction for comparison between theory and experiment, {\em i.e.} that  $x_{eff}=x_s-x_i$.  

\begin{figure}[h]
\hspace*{-1cm}\includegraphics[width=3.2in,angle=-90]{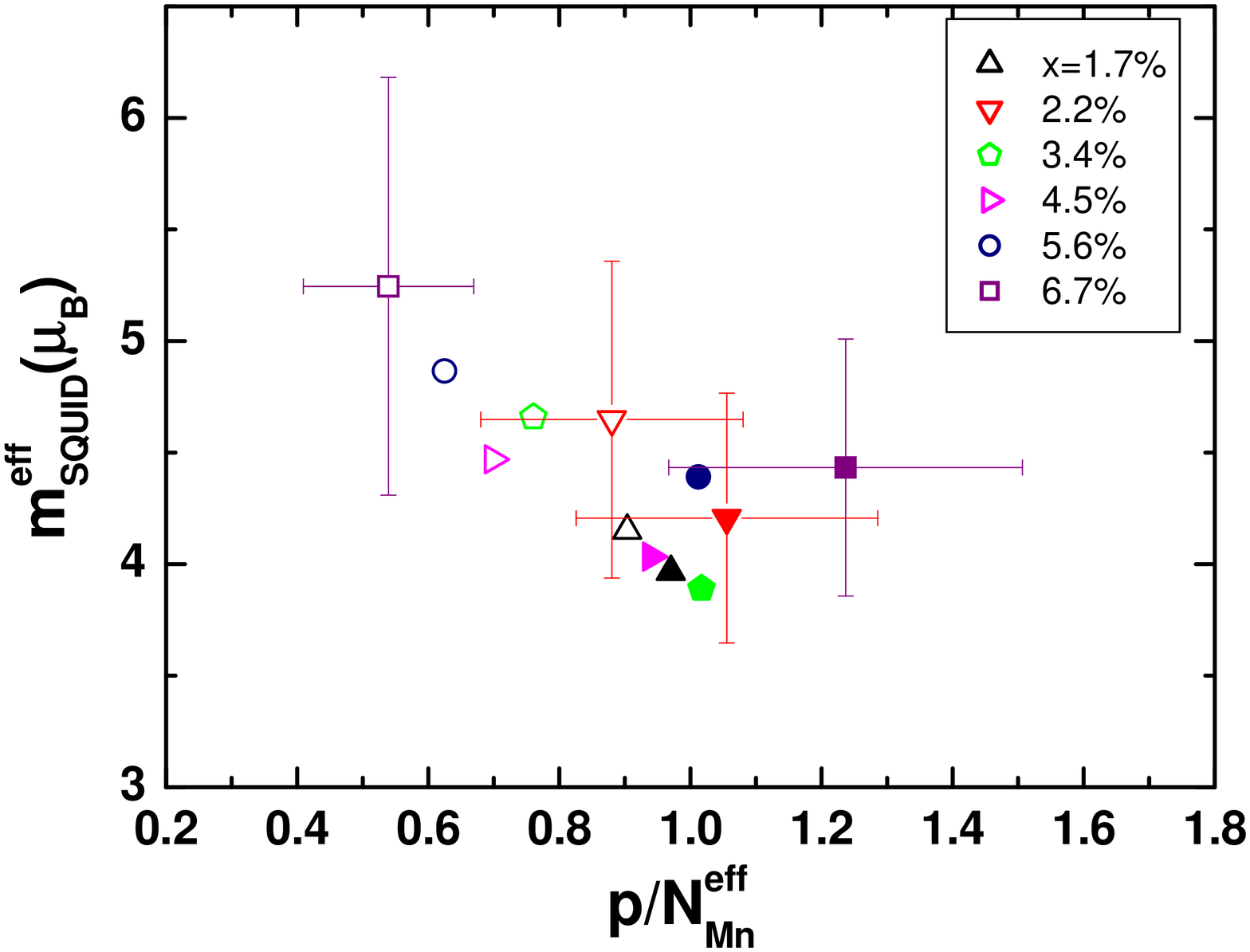}
\caption{SQUID magnetization per effective density of uncompensated
Mn$_{\rm Ga}$ local moments in as-grown (open symbols) and
annealed (filled symbols) (Ga,Mn)As materials plotted as a function of hole density per
effective density of uncompensated
Mn$_{\rm Ga}$ local moments.} 
\label{Meff-p}
\end{figure}

To obtain the individual Mn impurity concentrations 
we rely on Hall effect and magnetoresistance measurements at high magnetic field (up to ±16.5T) and low temperatures (down to 0.3K), from which we evaluate the experimental hole density $p$, 
after using a fitting procedure to separate normal and anomalous contributions to the Hall resistance.\cite{Edmonds:2003_a} We then assume that the single acceptors  Mn$_{\rm Ga}$ and double donors  Mn$_{\rm I}$ are the only impurities that
contribute to $p$, i.e., $p=(4/a_{lc}^3)(x_s-2x_i)$. From this expression and from
the total Mn concentration obtained by SIMS calibration ($x=x_s+x_i$) we can  estimate $x_{eff}$ for both as-grown and annealed samples. A detailed discussion of the uncertainties associated with this procedure is given 
elsewhere.\cite{Jungwirth:2005_b} The magnetic moment per effective Mn moment density, $m^{eff}_{SQUID}$, 
is shown in 
Fig.~\ref{Meff-p} as a function of $p/N_{Mn}^{eff}$ where $N_{Mn}^{eff}=(4x_{eff}/a_{lc}^3)$. 
In agreement with the predictions of the theory section,   $m^{eff}_{SQUID}$ falls 
within the range 4-5$\mu_B$ for all samples studied. Furthermore, although there 
is appreciable scatter, it can be seen that samples with lower hole densities tend to show higher
$m^{eff}_{SQUID}$, consistent with a negative contribution to magnetization from antiferromagnetically
coupled band holes.

\subsection{X-ray Magnetic Circular Dichroism}
XMCD measurements were performed using $(99\pm±1)$\% circular polarized x-rays from beamline ID8 at the European Synchrotron Radiation Facility. The samples are briefly etched in concentrated HCl prior to the measurements in order to remove Mn oxide rich surface layers, which may obscure the signal from the (Ga,Mn)As due to the relatively short probing depth of the measurement.\cite{Edmonds:2004_c} 
After etching, total electron yield and fluorescent yield measurements are in quantitative agreement, indicating a uniform distribution of Mn.

\begin{figure}[h]
\includegraphics[width=4.0in,angle=0]{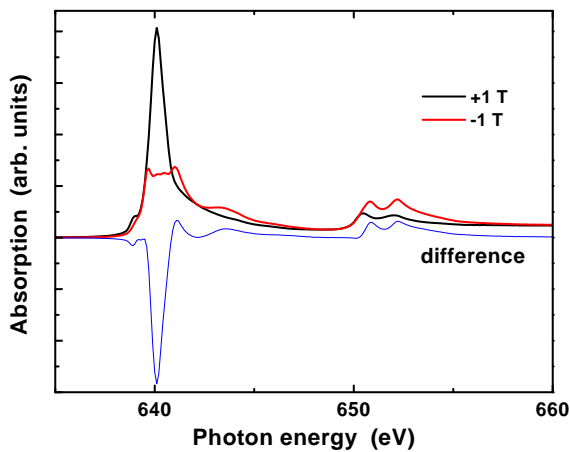}
\caption{Mn L$_{3,2}$ x-ray absorption spectra for an annealed (Ga,Mn)As sample with nominal Mn
doping 8.4\%, for parallel and antiparallel orientations of the external magnetic field and the x-ray helicity
vector.  The blue line shows the difference between the two spectra.} 
\label{xmcd}
\end{figure}

Fig.~\ref{xmcd} shows Mn L$_{3,2}$ x-ray absorption spectra for an annealed (Ga,Mn)As sample with $x=8.4$\%, for parallel and antiparallel orientations of the external magnetic field and the x-ray helicity vector. The sample temperature is 6K, and the external magnetic field is $\pm$1T, 
applied perpendicular to the surface of the sample. A very large change in the absorption is observed on reversing the external field, with an asymmetry (difference to sum ratio) of up to 55\% at 
the L$_3$ peak.
L$_{3,2}$ absorption corresponds to transitions from the 2$p$ core states to the unfilled 3d states, so the Mn L$_{3,2}$ spectra gives direct information on the polarization of the Mn 3$d$ band. Applying the XMCD sum rules to the spectra allows quantitative and separate determination of the Mn 3$d$ ground state orbital and spin magnetic moments.\cite{Carra:1993_a}
The moments are obtained on a per atom basis, without requiring separate measurement of the Mn concentration, by normalizing to summed absorption signal. There are, however, inherent uncertainties in the application of the sum rules, in particular due to mixing of the 2$p_{3/2}$ and 2$p_{1/2}$ states which prevents the separate integration over each of the spin-orbit split core levels. Comparison of calculated spectra with their corresponding ground state moments reveals that a correction factor of 1.47 is required for the spin moment to account for this mixing.
 The Mn 3$d$ moments obtained from XMCD are shown in Table~I, for two annealed samples with low
 and high Mn doping. In both cases, magnetic moments of around 4.5$\mu_B$ are obtained, 
 in agreement with the 
 SQUID results. Moreover, the measured moment and the property that it is independent of Mn doping is in very good agreement 
 with the calculated $d$-projected magnetic moment shown in the inset of Fig.~\ref{tba2}. 
 We note here that the calculations 
 shown in  Fig.~\ref{tba2} account only  for the spin angular momentum contribution to
 magnetization since SO-coupling effects were neglected in the TBA/CPA calculations.
 
 \begin{table}[h]

\begin{tabular}{cccc}
 \hline
$x$ &   $m^{spin}_{XMCD}$      & $m^{orb}_{XMCD}$        &   $m^{spin}_{XMCD}+ m^{orb}_{XMCD}$   \\

 (\%)& ($\pm0.3\mu_B$)     &  ($\pm0.03\mu_B$)   &  ($\pm0.3\mu_B$)  \\
\hline
2.2& 4.3  &  0.15        &  4.5  \\
\hline
8.4& 4.3 & 0.16 & 4.5 \\

\hline
\end{tabular}

\caption{Mn 3$d$ moments obtained from XMCD and decomposed into the spin and orbital contributions
in annealed samples with nominal Mn doping 2.2 and 8.4\%.}
 \label{table1}

\end{table}

\section{Conclusions}
We report on a combined theoretical and experimental analysis of the spontaneous magnetization in 
(Ga,Mn)As diluted magnetic semiconductor ferromagnets.  We find that the thermodynamic magnetization 
is dominated by a large local moment contribution of 5$\mu_B$ from nearly collinear substitutional 
Mn atoms.  Evaluation of the smaller magnetization contribution from the valence band system that couples the 
local moments together involves a number of subtleties.  In this paper we  included hole-hole exchange 
interactions and we also accounted for spin-orbit coupling which means that no valence band orbital is completely spin-polarized 
and which substantially changes the overall electronic structure.   Quantum fluctuations of the 
band and local moment orientations also play a role because of the antiferromagnetic interaction between
band and local moment spins.   The end result of all these corrections is a magnetization per 
magnetically active Mn ion that is suppressed from 5$\mu_B$ by $\sim 5-10\%$. 
Comparison with experimental data can be made reliably only after accounting for the 
formation of interstitial Mn complexes during the MBE growth, and for their subsequent 
removal by post-growth annealing.  Once these corrections have been applied, we find, within the
experimental error bars, 
agreement between theory and experiment.  The interpretation of XMCD magnetization measurements, which 
capture only the d-electron contribution, requires a recognition of the hybridized p-d character 
of both local moment and band-electron contributions to the magnetization.  Comparison of these 
measurements with TBA/CPA calculations provides experimental support for the applicability
of the kinetic-exchange model in (Ga,Mn)As ferromagnets.  Finally, our combined theoretical and 
experimental work demonstrates that non-collinearity does not play a significant role in the 
magnetization of high-quality metallic (Ga,Mn)As ferromagnets.

\label{conclusions}

\section*{Acknowledgment}
We thank Julio Cesare, Peter Bencok and Nick
Brookes for their contributions to the XMCD experiment. This work was supported by the Grant Agency
of the Czech Republic through Grant No.  202/05/0575, by the 
Academy of Sciences of the  Czech  Republic  through 
Institutional Support No. AV0Z10100521, by the  EU  FENIKS  project
EC:G5RD-CT-2001-00535, by the UK EPSRC through Grant GR/S81407/01,
by the Welch Foundation, and the US Department of Energy under grant DE-FG03-02ER45958.

\end{document}